# Modeling the Temperature Dependent Material Dispersion of Imidazolium Based Ionic Liquids in the VIS-NIR

**Yago Arosa, Bilal S. Algnamat, Carlos Damián Rodríguez, Elena López Lago, Luis Miguel Varela, Raúl de la Fuente**

*Grupo de Nanomateriais, Fotónica e Materia Branda. Departamentos de Física Aplicada e de Física de Partículas. Universidade de Santiago de Compostela. E-15782. Santiago de Compostela. Spain*

**Abstract.** A thorough analysis of the refractive index of eleven 1-alkyl-3-methylimidazolium based ionic liquids with three different anions, tetrafluoroborate $BF_4$, bis(trifluoromethylsulfonyl)imide $NTf_2$ and trifluoromethanesulfonate OTf, is reported. Refractive indices were estimated, in the temperature interval from 298.15 K to 323.15 K, using an Abbe refractometer to determine the value at the sodium D line and white light spectral interferometry to obtain dispersion in the range of wavelengths from 400 to 1000 nm. The first part of the manuscript is focused on the dependence of refractive index with wavelength, temperature, cation alkyl chain length and anion nature. Once the main features are detailed, and in order to explain the experimental trends, a model for the refractive index is considered where its square is expressed by a single resonance Sellmeier dispersion formula. This formula has two coefficients, the first one identifies the position of the resonance in the spectral axis, and the second one specifies its strength. It was found that, for a given compound, the resonance's position is independent of temperature, while the strength varies linearly with it. This model reproduces successfully the experimental data within the refractive index uncertainty. Furthermore, the model allows calculating the thermo-optic coefficient and its wavelength dependence.

1. **Introduction**

Ionic liquids (ILs) are molten salts that have melting points below 100 °C and are liquids in a wide temperature range, usually around and below room temperature. The interaction between anion and cation leads ILs to present interesting physical and chemical properties. They have a very low vapor pressure, negligible volatility, good chemical stability and high conductivity; they have wide electrochemical windows and are good solvents for a variety of organic and inorganic solutes [1]. Furthermore, due to the high numbers of IL-forming ions and their achievable combinations, ILs are highly tunable and can be tailored for specific applications. These include electrochemical applications, material synthesis and engineering, energy storage, process engineering, biotechnology, electronics or sensor technology. Moreover, thanks to the growing of optofluidics and related photonic areas, ILs have recently attracted some attention in the field of optics and photonics. To give some examples, ILs have been used for mineral and crystal analysis [2], to build variable focus lenses [3,4], to create IL-integrated-devices [5,6], to develop telescope liquid mirrors [7] or to design optical sensors [8,9], among others. However, the literature about optical properties of ILs is fairly scarce. The main optical property studied up to now is the refractive index. Studies are mainly devoted to its implication in process design, new materials development and the analysis of the correlation between the refractive index and other physical properties. With these objectives in mind, the great majority of works [10-13] were centered on the determination and analysis of the refractive index at a single wavelength, typically, the sodium D line (corresponding to a wavelength $\lambda$ = 589.3 nm). However, the potential use of any optical material depends on the accurate knowledge of its chromatic dispersion, that is, the variation of the refractive index with frequency or wavelength in the spectral range of interest. Furthermore, precise knowledge of chromatic dispersion is not only essential for the optical characterization of any material but it also can be used to obtain valuable information about various physical properties and chemical composition. Only some recent works have faced the measurement and analysis of dispersion. We highlight the work of Chiappe et al. [14] who experimentally determined the refractive index of 9 ILs at five wavelengths in the Vis and NIR spectral region and considered its temperature dependence. Besides, in a previous work [15] we measured the refractive index of 14 ILs over a quasi-continuum of wavelengths from 400 to 1000 nm at a constant temperature of 300.15 K. Both works centered their study on 1-alkyl-3-methylimidazolium based ILs, one of the best known families. Unlike other works, this paper is not focused on the correlation of refractive index and other physico-chemical properties but on the modeling of material dispersion and the analysis of the effect of temperature. We have also chosen for our study a group of ILs based on the imidazolium cation. The anions involved are $BF_4$, $NTf_2$ and $OTf$, making a total of 11 liquids. They are shown in fig. 1.

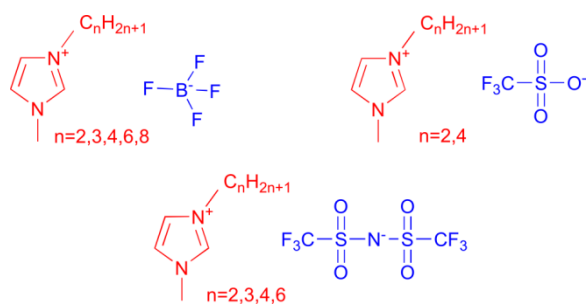

Figure 1. Chemical structure of the ILs studied in this work: [$C_n$MiM][$BF_4$], [$C_n$MiM][$NTf_2$] and [$C_n$MiM][$OTf$].

## 2. Experimental section

All compounds were supplied by IoLiTec. They were dried by stirring at room temperature in an evaporation flask connected to a vacuum pump to reduce pressure, during 48 h. After the drying process, they were kept in glass vials and closed with screw caps fitted with silicone septum to ensure a secure seal and preventing their contact with air moisture. Densities were measured with an Anton Paar DSA-5000M vibrating tube density and sound velocity meter at the same temperatures as the refractive indices with an uncertainty of 0.01 K. The apparatus was calibrated by measuring the density of bidistilled water and dry air at atmospheric pressure according to the manual instructions. The overall precision in experimental density measurements for all samples was found to be better than ±2 x $10^{-6}$ g·$cm^{-3}$.

We have used white light spectral interferometry (WLSI) to measure the refractive index of the ILs listed in Fig. 1 in the range from 400 to 1000 nm at six temperatures from 298.15 K to 323.15 K, at every 5 K. As it is shown in Fig. 2, the apparatus comprises a white light source, a Michelson interferometer and a spectrometer. Both, the interferometer and the spectrometer are homemade. A detailed explanation of the experimental setup can be found in [16]. A quartz cell, Hellma 100-QS 1 mm, with an IL sample is disposed in one of the arms of the interferometer and an external thermostatic bath keeps constant the temperature of the sample up to 0.1 K. The interference of the beams that travel in the interferometer arms is spectrally decomposed and detected in a prism spectrometer. The phase difference between beams is extracted from the interferogram. After subtraction of the optical path introduced by the quartz cell, the following phase is obtained [16]:

$$\varphi(\lambda) = \frac{4\pi}{\lambda}\left[d(n(\lambda)-1) - L\right] - 2k\pi, \tag{1}$$

where $\lambda$ is the wavelength, *d* is the sample thickness, *n* is the refractive index, *L* is the path difference in air in the interferometer arms and *k* is the interference order corresponding to the last maximum (in this way, the condition of maximum of interference can be written as $\varphi = 2q\pi$ with $q = 1, 2, \ldots, k$). The value of *d* is measured with a high precision micrometer Mitutoyo MDH-25M. *L* is obtained with a 0.5 m focal length commercial diffraction grating spectrometer (SpectraPro 500, not shown in the figure), by applying also WLSI [15-16]. On the other hand, to determine *k,* we measured the refractive index at the sodium D line with a commercial Abbe refractometer, Atago DR-M2, with a temperature controller (resolution 0.1 K). The refractometer was calibrated with deionized water and its resolution at the sodium D line was $1 \times 10^{-4}$. Once *d*, *L*, and *k* are solved, dispersion of the refractive index is directly retrieved with eq. (1). Our apparatus has been tested to retrieve refractive indices with accuracy better than $3 \times 10^{-4}$ [17].

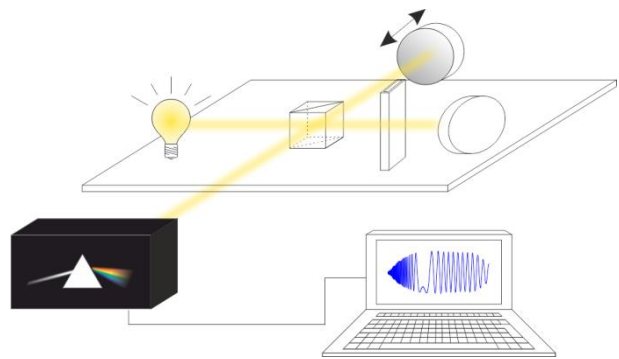

Figure 2. Experimental setup. A white light source illuminates a Michelson interferometer with the sample in one of its arms. The light at the output spectrometer is spectrally decomposed in a prism spectrometer and analyzed with a laptop computer.

## 3. Results and discussion

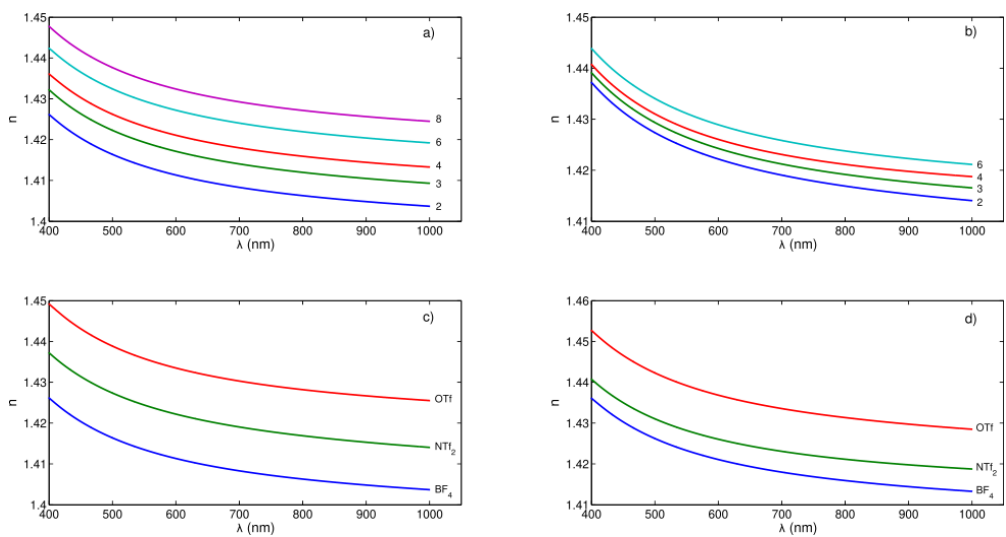

Figure 3. Material dispersion of refractive index at T = 298.15 K. Figures (a) - (d) corresponds to groups [$C_n$MiM][$BF_4$] (I), [$C_n$MiM][$NTf_2$] (II), [$C_2$MiM][X] (III) and [$C_4$MiM][X] (IV). Being n the carbon number in a) and b) and X representing different anions in c) and d).

Fig. 3 shows the material dispersion for the 11 ILs at a temperature of 298.15 K. Measured values of refractive index at selected Fraunhofer lines and temperatures are provided in the supporting information (Table S1). For clarity, we have arranged the ILs into four groups. In the first group (I), the compounds share the *BF$_4$* anion while the cations differ in the number of carbons on the alkyl chain. The second group (II) has similar characteristics but the common anion is *NTf$_2$*. In the third group (III) we consider compounds with [*C$_2$MiM*] cation and different anions, while in the fourth group (IV) the common cation is [*C$_4$MiM*]. There are some distinguishing features that must be highlighted. First, in the measured spectral range, the refractive index curves follow the typical behavior of the normal dispersion regime; that is, the refractive index decreases as wavelength increases and the dispersion is higher for smaller wavelengths. Moreover, in spite of being the refractive indices quite similar, all curves are well separated and present similar chromatic dispersion as they are mostly parallel. That means that the analysis of the refractive index at one given wavelength as a function of the carbon chain or of the temperature grossly shows general features that can be extrapolated to other wavelengths. For a widespread study at a single wavelength without considering dispersion we refer the reader to previous works that considered the refractive index at the sodium D line [10-13]. Considering the first two groups (I and II) where imidazolium cations with different alkyl length share the same anion, it is observed that the refractive index increases with the number of carbons in the alkyl tail of the cation. This is also true for the two compounds that share the *OTf* anion. However, the molar mass increases with the number of carbons while density decreases, just the inverse as the refractive index does. That means that variation of molar refraction (or mean polarizability) with the number of carbon must be responsible for this behavior (see below). Furthermore, for ILs with the same alkyl chain length, the refractive index is greater for the compounds with *NTf$_2$* anion. Otherwise, the refractive index changes in larger extent with the number of carbon atoms for the *BF$_4$* group, while, for the *NTf$_2$* group it is more packed. This behavior is related to the relative weight variation in the compounds as carbons in the alkyl chain are added. The *NTf$_2$* anion has larger polarizability and occupies more volume than the *BF$_4$* anion (see Fig. 4 below) and, therefore, the addition of new carbons in the alkyl chain has a relative lower impact for *NTf$_2$* ILs than for *BF$_4$* ILs. As a consequence, the variation of refractive index for each added carbon in the alkyl chain is higher for the *BF$_4$* family than for the *NTf$_2$* family.

On the other hand, groups III and IV are very similar. ILs with the *BF$_4$* anion have the lowest refractive index, followed by ILs with *NTf$_2$* anion, while ILs with *OTf* anion are the ones with the greatest refractive index. ILs in group IV have refractive index values close to those in group III. The behavior of the refractive index curves at other temperatures share basically the same characteristics. Regarding the refractive index dependence on temperature, it decreases as the temperature rises. However, the studied ILs present a relatively small thermo-optic coefficient as shown below in the text. Thus, temperature has a reduced impact on the refractive index and dispersion dominates its behavior. In consequence, the refractive index curves maintain their shapes despite their values are modified as temperature changes.

Coming back to the relation between refractive index and length of the alkyl tail of the cation, the change is not uniform, but it varies slowly as the length increases. Some authors [18,19], have already pointed that it is the polarizability rather than the refractive index that linearly varies with the length of the alkyl chain. The polarizability is related to the response of a material to an external electric field of a given frequency. At optical frequencies, it provides a value of how strong the material interacts with light. The molar refraction, $R$, of a given compound is a magnitude that describes the electronic polarizability of a mol of such compound, and can be calculated using the Lorentz-Lorenz equation [10,11,15,20]:

$$R = V_m \frac{n^2 - 1}{n^2 + 2}, \quad (2)$$

Where, $V_m = M/\rho$ is the molar volume, being $M$ the molar mass and $\rho$ the density of the compound. To obtain the values of molar refractions we have also measured the density of the 11 ILs. The density values at different temperatures, as well as the coefficients of a linear fitting over temperature, are accessible in the supporting information (Table S2 and S3).

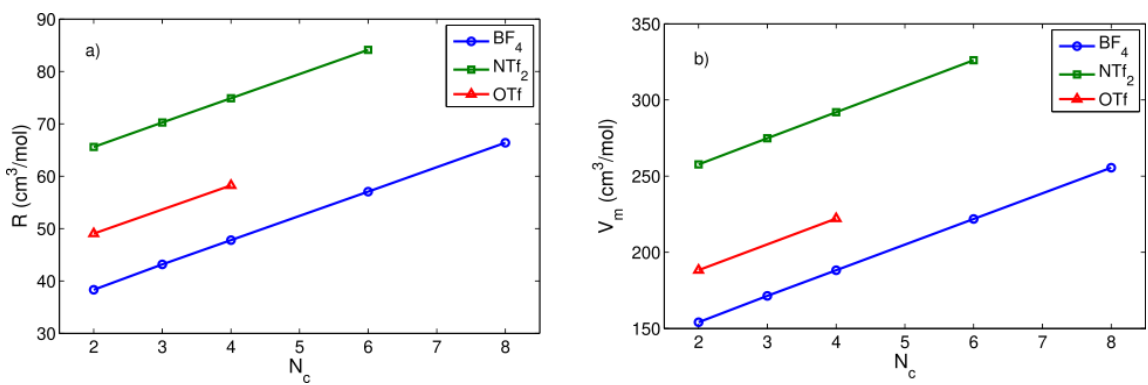

Figure 4. (a) Molar refraction at sodium D line for different anions and T = 298.15 K as a function of the number of carbon atoms in the alkyl chain of the IL cation; (b) Molar volume as a function of alkyl chain for a T = 298.15 K.

In Fig. 4(a) and in the supporting information (Table S4) we show the highly linear relation between molar refraction and number of carbons in the cation chain $R = R_0 + \Delta R N_c$, for the sodium D line at a temperature of 298.15 K. We note that molar refraction of the different ILs is ordered in the same way as the refractive index. The slope of the molar refraction changes little for the three studied groups. These changes mainly seem to be due to the increase in molar mass when adding a methyl group. The intercept value is related to the combination of the polarizability of the imidazolium ring and the anions as well as their weight. As the imidazolium ring is the same in all the compounds, the main differences in molar refraction are related to the anions. Heavy and easily polarizable anions such as *NTF$_2$* produce larger intercept values than light and less polarizable anions such as *BF$_4$*.

As already noted by other authors [10], the molar volume is almost a linear function of cation alkyl length, $V_m = V_0 + \Delta V N_c$. This linear increasing of molar volume with the alkyl length of the cation can be interpreted as an elongation of the molecule in a given direction, and subsequently, the change in molar volume is proportional to this elongation. That means that the molecule takes the form of a spheroid or an ellipsoid rather than a sphere. Indeed, *ΔV* is almost the same for families with different anions. From our measurements, a value for *ΔV of* 17 cm³/mol was obtained, in coincidence with the value reported in [10]. It is the molar volume of the *CH$_2$* group and it increases slightly with temperature since density decreases. Fig. 4(b) shows the molar volumes of ILs with *BF$_4$*, *NTf$_2$* and *OTf* anions as a function of alkyl chain length at a temperature of *T* = 298.15 K. Molar masses and coefficients of the linear fits are accessible in the supporting information (Tables S5 and S6).

The eq. (2) can be rearranged to be expressed in terms of the molar refractivity and the free molar volume $f_m$:

$$n^2 - 1 = \frac{3R}{V_m - R} = \frac{3R}{f_m}, \qquad (3)$$

This free molar volume is the difference between the total available molar volume and the molar refractivity volume. It is a measurement of the molar volume which does not interact with light. Eq. (2) and thus eq. (3) have to be carefully handled. As refractive index is inversely proportional to free molar volume, it may seem that the refractive index is directly proportional to the density. This behavior has been previously suggested [21] based on the identification of free volume with a measurement of the degree of molecular package. High molecular packages are produced by high densities and thus low free molar volumes. On the opposite case, low molecular packages mean low densities and high free molar volumes. This framework correctly explains processes in which molar refractivity does not change. For instance, the decreasing of refractive index with temperature: as temperature increases, density decreases so free molar volume increases and the refractive index diminishes. However, this vision is not complete and cannot explain the experimental behavior of some ionic liquid families when alkyl chain length is considered, as noted in [10]. The whole picture about free molar volume includes not only the degree of molecular packaging but also the molar refractivity. When increasing the alkyl chain inside a specific ionic liquid family not only the molecular volume is changed but also the molar refractivity. As previously was shown, both magnitudes linearly evolve with alkyl chain. However, these magnitudes do not change in the same amount as alkyl chain increases, that is, both magnitudes present different slopes. It is the relationship between these two magnitudes which determines the evolution of refractive index with alkyl chain as described in [20]. Taking all this considerations into account, equations (2) and (3) can be used without further risk.

As both, molar refractivity R and molar volume $V_m$ are linear functions; the eq. 3 is a rational function of the carbon number for liquids within the same family. We introduced the linear fittings for the molar volume and molar refractivity in such equation. We were able to correctly reproduce the experimentally measured refractive indices, testing the validity of the Lorentz-Lorenz equation for ionic liquids.

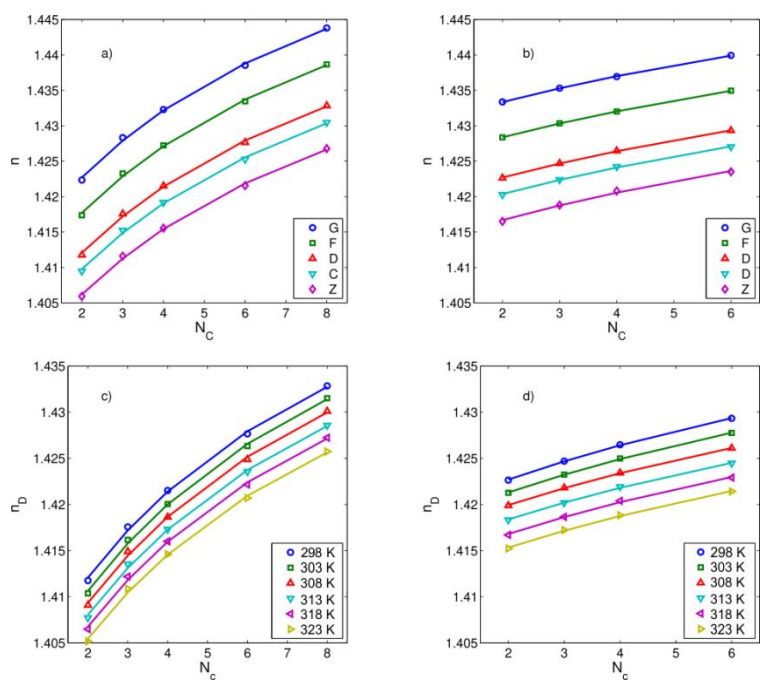

Figure 5: (a-b) Refractive indices as a function of the alkyl chain respectively for groups I ($BF_4$) and II ($NTF_2$) at different Fraunhofer lines: G ($\lambda$ = 430.1 nm), F ($\lambda$ = 486.1 nm), D ($\lambda$ = 589.6 nm), C ($\lambda$ = 656.3 nm) and Z ($\lambda$ = 822.7 nm); (c-d) Refractive index at the sodium D line as a function of alkyl chain length at different temperatures respectively for groups I and II.

In Fig. 5 (a-d) the fits for compounds with *BF*$_4$ and *NTf*$_2$ anions at some wavelengths and temperatures are shown. The difference between experimental and fitted data is always less than 4.4 x 10$^{-4}$ and less than 2.7 x 10$^{-4}$ for ILs sharing the *BF*$_4$ and *NTf*$_2$ anions, respectively. Figs. 5(a-b) show the behavior of the refractive index with the number of carbon atoms in the cation alkyl tail at constant temperature T = 298.15 K and different wavelengths. The curves at different wavelengths present similar behavior but different values. For the *NTf*$_2$ anion, the change in refractive index as a consequence of the increasing alkyl chain is smaller than for the *BF*$_4$ anion. In addition, the relation is more markedly linear for the family with *NTf*$_2$ anion. In Fig. 5(c-d) we compare the variation of refractive index for different cations at different temperatures. The shown temperatures increase every 5 K, and in this case the curves are evenly spaced, an evidence of the linear relation of refractive index and temperature. As commented above, the contribution of each *CH*$_2$ group to the molar refraction and molar volume is almost independent of temperature and the nature of the anion. So, the differences in the refractive indices are mainly due to the intercepts $R_0$ and $V_0$.

In fact, with respect to the change in refractive index with temperature it is seen that it decreases almost linearly for all compounds. Fig. 6 shows some examples. The slope, which is known as thermo-optic coefficient (TOC) is always negative and, in absolute terms, it increases with the cation alkyl chain length. In absolute terms too, the TOC is smaller for compounds that share the *BF*$_4$ anion and greater for compounds sharing the *NTf*$_2$ anion. In all cases the absolute value of the TOC is very small (at it is usual), less than 1 x 10$^{-3}$ K$^{-1}$. Regarding the dependence on wavelength, the TOC decreases in absolute terms, approaching zero as the wavelength increases.

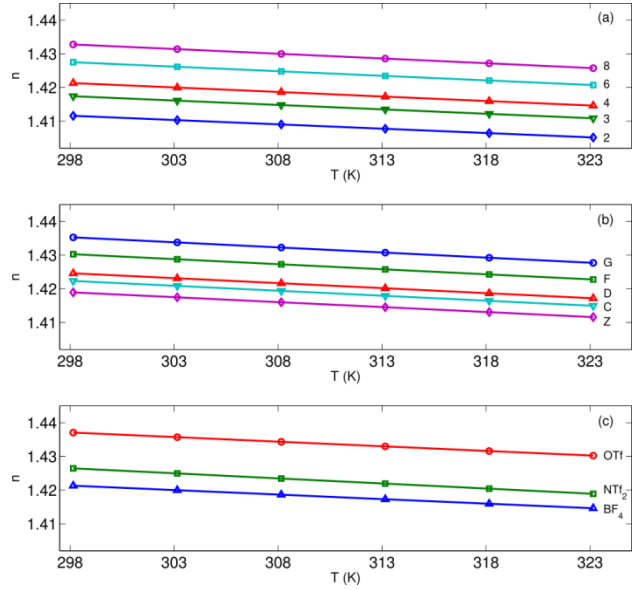

Figure 6. Refractive indices as a function of temperature. (a) [C$_n$MiM][BF$_4$] at the Sodium D line; (b) [C$_3$MiM][NTf$_2$] for different Fraunhofer lines; c) compounds which share the [C$_4$MiM] cation at the sodium D line.

In order to model the dependence of refractive index on wavelength and temperature, we tried to fit the experimental data with a temperature dependent dispersion equation. The widespread equations used to represent the refractive index of a material are the Cauchy and the Sellmeier dispersion formulas [22]. While the Cauchy equation is simpler (it merely corresponds to an expansion in power $\lambda^{2n}$, with $n$ integer), the Sellmeier equation can provide more information about the liquid. Indeed, the Cauchy formula is nothing but an approximation of the Sellmeier when the wavelength is far from singular points. The general expression of Sellmeier equation is:

$$n^2 - 1 = \sum_{i=1}^{N} \frac{c_i \lambda^2}{\lambda^2 - \lambda_i^2}, \qquad (4)$$

Where, the constants $c_i$ and $\lambda_i$ are related with the absorption spectra of the compound. $c_i$ relates with the strength of an absorption peak and $\lambda_i$ with its position. So, the Sellmeier equation not only provides information about refractive index, but also about absorption. When using this equation to describe the refractive index over a wide spectral range, it is common to employ three resonances with two of them located in the UV and the other one in the IR. Typically, the ultraviolet resonances are below 200 nm and the infrared resonances are above 3 μm. The proximity of the ultraviolet resonances to the visible range is responsible for the normal dispersion of the refractive index curve. Also, the curvature of refractive indices as a function of wavelength is convex for our set of liquids, which means that the ultraviolet resonances dominate over the infrared ones. Taking this into account and trying to reduce the number of resonances of our fit, we found that a Sellmeier formula with a single UV resonance performs remarkably well. Initially we consider independent fits for the same compound at different temperatures. Soon we realized that the value of the resonant wavelength seems to be independent of temperature, while the strength of the resonance varies almost linearly with temperature. Therefore, we consider the fit:

$$n^2(\lambda, T) - 1 = [c_1 + c_2 \Delta T] \frac{\lambda^2}{\lambda^2 - \lambda_{uv}^2}, \tag{5}$$

Where, $\Delta T = T - T_0$, being $T_0$ the temperature of reference (in this work we take $T_0$ = 310.65 K, the center of the measured temperature interval). Some features about this fit must be highlighted. The relation discussed above between refractive index and alkyl chain length for ILs sharing the same anion suggests that the resonance does not change with the cation alkyl chain length. Consequently, we fix the wavelength of the resonance while changing the cation alkyl chain length. Of course, this worsens somewhat the fit, but keeps small the root mean square deviation. Once the resonance is determined, we calculate the coefficient $c_1$ and $c_2$ in eq. (5) by applying linear least squares method.

Besides, since our spectrometer intentionally contains a prism as the dispersive element, the relation between pixel position in the camera and wavelength is highly nonlinear. This means that the spectrum is oversampled at the smaller wavelengths and, in turn, a direct fit will perform best in the visible range than in the IR range. Therefore, we resampled the data picking quasi evenly spaced wavelengths to obtain a regular fit in the whole spectral range.

The coefficients for the 11 ILs are listed in Table 1. The temperature independent term in eq. (2) is the dominant one. For any wavelength and temperature, the remaining term is more than one thousand times smaller. Thus, denoting the refractive index squared at $T_0$ as $n_0^2 = 1 + c_1 \lambda^2 / (\lambda^2 - \lambda_{uv}^2)$ and $\Delta n^2 = n^2 - n_0^2 = c_2 \Delta T \lambda^2 / (\lambda^2 - \lambda_{uv}^2)$, we have $\left|\Delta n^2\right|/n_0^2 \leq 0.001$ for every $\lambda$ and $T$. Regarding the coefficients, one main characteristic of $c_1$ and $c_2$ is that they increase in absolute value with the alkyl chain length; so, not only refractive indices but absolute TOCs increase with the number of carbon atoms in the cation chain. Meanwhile, the resonant wavelength barely changes. Furthermore, the coefficient $c_2$ is always negative in accordance with the reduction of refractive index with increasing temperature. Finally, regarding components sharing the same cation, the order of increment for the coefficients $c_1$ and $c_2$ is $BF_4$ < $NTf_2$ < $OTf$ and $BF_4$ <$OTf$ < $NTf_2$, respectively.

Table 1. Sellmeier coefficients of the refractive index of each ionic liquid (it is assumed that the wavelength unit is microns).

| IL | $\lambda_{uv}^2$ [$10^{-2}$ µm$^2$] | $c_1$ | $c_2$ [$10^{-4}$ K$^{-1}$] |
|---|---|---|---|
| [C$_2$MiM][BF$_4$] |  | 0.9515 (62) | -7.0100 (53) |
| [C$_3$MiM][BF$_4$] |  | 0.9672 (69) | -7.1794 (65) |
| [C$_4$MiM][BF$_4$] | 1.13260 (94) | 0.9777 (67) | -7.3547 (61) |
| [C$_6$MiM][BF$_4$] |  | 0.9946 (73) | -7.4826 (74) |
| [C$_8$MiM][BF$_4$] |  | 1.0088 (66) | -7.7626 (59) |
|  |  |  |  |
| [C$_2$MiM][NTf$_2$] |  | 0.9811 (95) | -8.049 (12) |
| [C$_3$MiM][NTf$_2$] | 1.10095 (90) | 0.9867 (67) | -8.1557 (61) |
| [C$_4$MiM][NTf$_2$] |  | 0.9916 (77) | -8.3092 (82) |
| [C$_6$MiM][NTf$_2$] |  | 0.9988 (66) | -8.7014 (60) |
|  |  |  |  |
| [C$_2$MiM][OTf] | 1.14500 (13) | 1.0117 (69) | -7.7013 (65) |
| [C$_4$MiM][OTf] |  | 1.0206 (78) | -7.5887 (83) |

We must emphasize that we have tried to perform other types of fits but they were not consistent. For instance, we considered a Sellmeier formula with two resonances, one in the UV and the other in the IR. Each fit performed well, but the behavior of the IR resonance for a given compound at different temperatures was erratic. We also considered another fit with a constant term added to eq. (5). The value obtained for the constant did not have sense in many cases and the dependence on *T* of the strength of the resonance was not clear. In contrast the simple fit of eq. (5) is very regular. In fact, it allowed us detecting some errors in the measurement of one compound ([*C$_3$MiM*][*NTf$_2$*]). After repeating the measurement the result was optimal.

The temperature dependence of the refractive index can be determined by the T-derivative of eq. (5). Assuming that $c_i$ and $\lambda_{uv}$ are independent of temperature, it is obtained as:

$$\frac{dn}{dT} = \frac{c_2}{2n} \frac{\lambda^2}{\lambda^2 - \lambda_{uv}^2}. \tag{6}$$

At first sight, and since the refractive index varies with temperature, the TOC does. However, we can approximate $n \approx n_0$ in the denominator of eq. (6) and thus, on this level of approximation, the TOC varies only with wavelength. That means:

$$\frac{dn}{dT} \approx \frac{1}{2n_0 \Delta T} \Delta n^2(\lambda, T). \tag{7}$$

We see that the wavelength dependence of the TOC also follows a single resonance Sellmeier dispersion formula with the peculiarity that the 'strength' is negative. With the same level of approximation, the refractive index verifies:

$$n(\lambda, T) \approx n_0(\lambda) + \frac{\Delta n^2(\lambda, T)}{2n_0(\lambda)}, \tag{8}$$

That is also linear with temperature. Examples of the goodness of this fit are shown in Fig. 7. For the three plotted compounds at each temperature the mean square deviation between experimental and fitted data is less than 3.5 x $10^{-5}$, one order of magnitude smaller than the experimental uncertainty.

In Fig. 8, the dispersion of TOC for the different compounds is plotted. The most remarkable feature is that, as the refractive index does, the dispersion of TOC is larger for smaller wavelengths, and it tends to a constant value $c_2/2\sqrt{1+c_1}$ when $\lambda \gg \lambda_{uv}$. Besides, the absolute value of TOC increases with the length of alkyl chain. As noted in [23], this can result from higher dilatation coefficients associated with lower degrees of hydrogen bonding. With respect to compounds sharing the same cation, the order change with respect to Fig. 3. In this case, the order from the smallest to the greatest TOC is $BF_4 < OTf < NTf_2$. This could be attributed to a change in the thermal dilatation that follows the same trend. Finally, we highlight that the TOC dispersion managed in this work disagree with the analysis of other authors who considered that the thermo-optic coefficient is independent of wavelength [14, 24]. Being the change of TOC small, less than 4% in the wavelength interval, it is greater than the variation of refractive index in the same interval which is less than 2%.

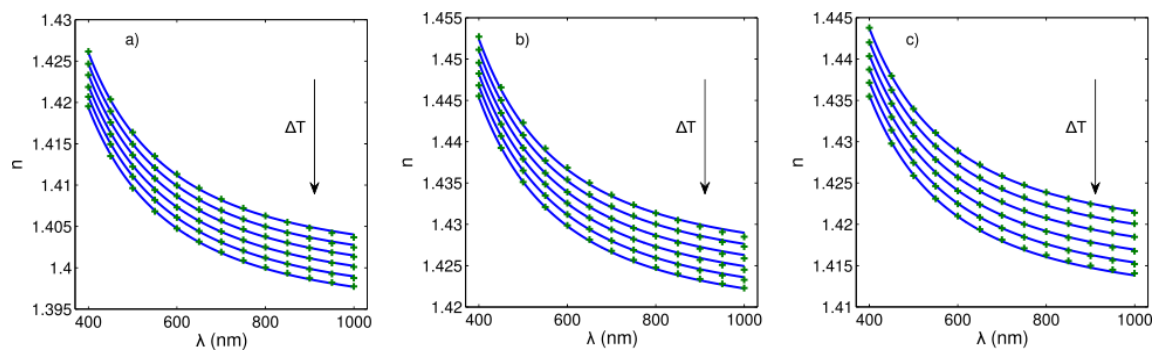

Figure 7. Experimental refractive index versus wavelength at different temperatures: experimental data (marks); values given by eq. (8) (lines). (a) $C_2MiMBF_4$; (b) $C_4MiMOTf$; (c) $C_6MiMNTf_2$. The temperature decreases vertically and goes from 298.15 K to 323.15 K every 5 degrees. For clarity, there are plotted experimental refractive indices at only some given wavelengths.

As the refractive index, TOC is also related with the variation of density and molar refraction with temperature. Differentiating eq. (2) it is straightforward to obtain the following relation:

$$\frac{1}{\rho}\frac{d\rho}{dT} + \frac{1}{R}\frac{dR}{dT} = \frac{d}{dT}\ln\left(\frac{n^2-1}{n^2+2}\right) = \frac{6n}{(n^2-1)(n^2+2)}\frac{dn}{dT}. \quad (9)$$

In this equation, the first term corresponds to minus the coefficient of thermal expansion and the second term to the temperature coefficient of molar refraction (or electronic polarizability). The meaning of this equation is that the magnitude of the TOC is the result of a delicate balance between the thermal expansion or electrostriction, and the thermal dependence of the electronic polarizability. Positive thermal expansion coefficient leads to a decrease of refractive index with temperature (i.e. negative TOC) while positive temperature coefficient of molar refraction leads to an increasing of refractive index with temperature. This last contribution can be calculated as a function of wavelength from the fitted refractive indices and densities. It turns out than it is a decreasing function of wavelength which varies in the range $(0.2 - 2.5) \times 10^{-5}$, considering all compounds and temperatures. Meanwhile, the thermal expansion coefficient is about $(5.8 - 6.8) \times 10^{-4}$. Therefore, the first contribution dominates the second one in eq. (9) as the relation between coefficients for every temperature and wavelength ranges from 0.3 to 7.6%. As a result, the refractive index decreases with temperature as it was discussed and as it is typical in ILs and other liquids. However, this result differs from that in [14] that deals with the temperature dependent dispersion of imidazolium-based ionic liquids with a phosphorus-containing anion. In that manuscript, the magnitude of the thermal dependence of the electronic polarizability was found to be smaller than, but comparable to, the magnitude of the thermal expansion coefficient (from 30% to 60%).

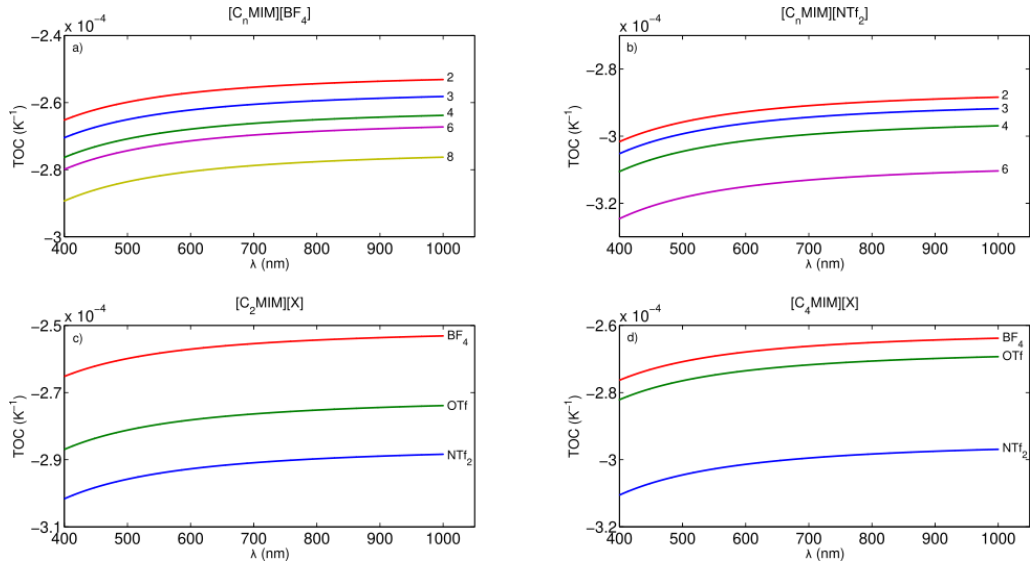

Figure 8. Dispersion of the TOC.

Finally, we come back to the variation of refractive index with alkyl chain length. Eq. (3) predicts that the square of refractive index is a linear rational function of the number of carbon atoms in the cation chain, $N_c$. The fit of refractive index given by eq. (5) will be consistent with this behavior if the coefficients $c_1$, $c_2$ are also linear rational function of $N_c$ with the same denominator. We performed this type of fit and the results are shown in fig. 9. First, we consider the fit for coefficient $c_1$ and then we performed the fit of coefficient $c_2$, fixing it to the same denominator as $c_1$. That means that the numerator must be fitted to a linear function of $N_c$.

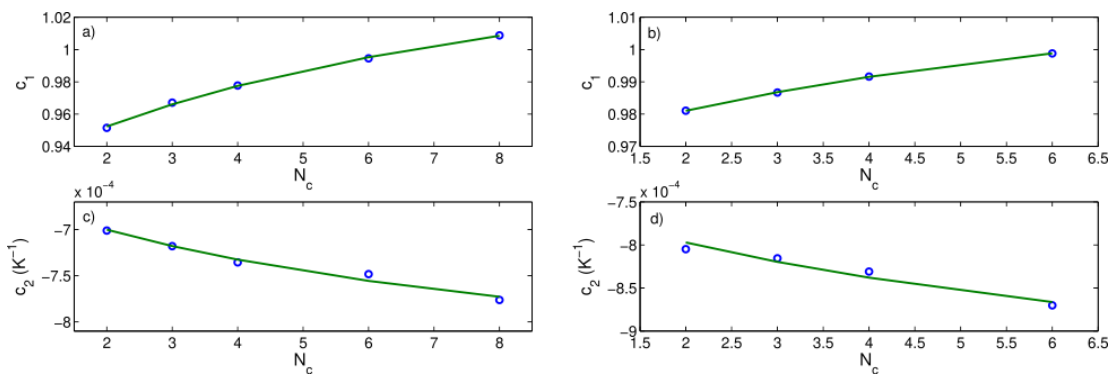

Figure 9. Value of the coefficients $c_1$, $c_2$ as a function of carbon number together with a fit to a linear rational function. The coefficients relate to (a,c) group I and (b,d) group II.

## 4. Conclusions

To summarize, the temperature dependent dispersion of 11 imidazolium based ILs was measured by WLSI in the spectral range from 400 to 1000 nm and a temperature interval from 298.15 K to 323.15 K, every 5 K. It was found that the refractive index curve for each ILs and temperature shows the characteristics of the normal dispersion regime, with larger refractive indices and refractive index variations at smaller wavelengths. Furthermore, dispersion is similar for all the ILs analyzed being its magnitudes the main difference. Considering ILs sharing the same anion, and taking into account the almost linear relation of molar refraction and molar volume with alkyl length, it was found that the refractive index squared can be modeled with a rational function of degree one. With respect to the relation between refractive index and temperature, it is almost linear with negative slope and a value which increases in absolute terms with alkyl chain length and decreases with wavelength. Besides, the variation of refractive index with the nature of the anion was also analyzed.

The refractive index squared of all the ILs studied in this work is modeled as a function of wavelength and temperature by a Sellmeier dispersion equation with a single UV resonance. The spectral position of this UV resonance is independent of temperature for a given IL but its strength varies linearly with it. Furthermore, we considered a resonance peak that remains constant while the length of the cation alkyl chain varies. This model was employed to calculate and analyze the TOC of the studied ILs which, with a good approximation, is independent of temperature. Refractive indices obtained with this model are in very good agreement with experimental data with deviations of the order of the refractive index uncertainty. Finally, the main thermodynamic contributions to the behavior of the TOC were pointed out: thermal expansion and the thermal dependence of molar refraction. It was found that the main contribution to the TOC is thermal expansion which explains why the refractive index of the studied ILs decreases as temperature is increased.

**Supporting Information description**

Table S1. Experimental refractive indices at different temperatures and wavelengths.
Table S2. Density (g/cm$^3$) as a function temperature as a function of temperature
Table S3. Coefficients of the linear fit of densities as a function of temperature
Table S4. Molar refraction coefficients for different wavelengths and temperatures ($R = R_0 + \Delta RN_c$)
Table S5. Molar mass of the studied ILs
Table S6. Molar volume coefficients for different temperatures ($V_m = V_0 + \Delta VN_c$)

**Acknowledgement**

Ministerio de Economía y Competitividad (MINECO) (MAT2014-57943-C3-1-P, MAT2014-57943-C3-2-P, MAT2017-89239-C2-1-P); Xunta de Galicia and FEDER (AGRUP2015/11, GRC ED431C 2016/001, ED431D 2017/06, ED431E 2018/08), C. D. R. F. thanks the support of Xunta de Galicia through the grant ED481A-2018/032.

Table of Contents

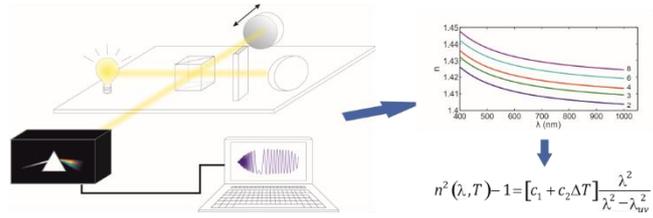